\documentclass{aa}  

\usepackage{graphicx}
\usepackage{txfonts}
\usepackage{mathrsfs}
\usepackage[colorlinks,allcolors=blue]{hyperref}
\usepackage{breakurl}
\usepackage{subfigure}

\def \src {\mbox{AX\,J0049.4$-$7323}}

\def \sw  {\emph{Swift}}
\begin{document}

   \title{X-ray and optical monitoring of the December 2017 outburst of the Be/X-ray binary  AX\,J0049.4$-$7323} 
   \titlerunning{X-ray and optical monitoring of the December 2017 outburst of AX\,J0049.4$-$7323} 

   \author{L. Ducci
          \inst{1,2}
          \and
          C. Malacaria\inst{3,4}
          \and
          P. Romano\inst{5}
          \and
          L. Ji\inst{1}
          \and
          E. Bozzo\inst{2}
          \and
          I. Saathoff\inst{1}
          \and
          A. Santangelo\inst{1}
          \and
          A. Udalski\inst{6}
          }

   \institute{Institut f\"ur Astronomie und Astrophysik, Kepler Center for Astro and Particle Physics, Eberhard Karls Universit\"at, 
              Sand 1, 72076 T\"ubingen, Germany\\
              \email{ducci@astro.uni-tuebingen.de}
              \and
              ISDC Data Center for Astrophysics, Universit\'e de Gen\`eve, 16 chemin d'\'Ecogia, 1290 Versoix, Switzerland
              \and
              NASA Marshall Space Flight Center, NSSTC, 320 Sparkman Drive, Huntsville, AL 35805, USA
              \and
              Universities Space Research Association, NSSTC, 320 Sparkman Drive, Huntsville, AL 35805, USA
              \and
              INAF -- Osservatorio Astronomico di Brera, via Bianchi 46, 23807 Merate (LC), Italy
              \and
              Warsaw University Observatory, Al. Ujazdowskie 4, PL-00-478 Warszawa, Poland
             }

   \date{Received ...; accepted ...}

 
  \abstract
   {\src\ (SXP\,756) is a Be/X-ray binary that shows an unusual and poorly understood optical 
   variability that consists of periodic and bright optical outbursts, simultaneous
   with X-ray outbursts, characterised by a highly asymmetric profile.
   The periodicity of the outbursts is thought to correspond to the orbital period
   of the neutron star.
   To understand the peculiar behaviour shown by this source, we performed the first 
   multi-wavelength monitoring campaign during the periastron passage of December 2017.
   The monitoring lasted for about 37 days and consisted of X-ray, near-ultraviolet,
   and optical data from the Neil Gehrels Swift Observatory, the optical $I$ band 
   from the OGLE survey, and spectroscopic observations of the H$\alpha$ line performed
   with the 3.9\,m Anglo-Australian Telescope.
   These observations revealed \src\ during an anomalous outburst having remarkably different properties compared to the previous ones.
   In the $I$ band, it showed a longer rise timescale ($\sim 60$ days 
   instead of 1--5 days) and a longer decay timescale.
   At the peak of the outburst, it showed a sudden increase in luminosity in the 
   $I$ band, corresponding to the onset of the X-ray outburst.
   The monitoring of the H$\alpha$ emission line showed a fast and highly variable
    profile composed of three peaks with variable reciprocal brightness.
   To our knowledge, this is the second observation of a variable three-peak H$\alpha$ profile  of a Be/X-ray binary, after A0535+26.
   We interpreted these results as 
   a circumstellar disc warped by tidal interactions with the neutron star 
   in a high eccentricity orbit during its periastron passage. 
   The fast jump in optical luminosity at the peak of the
   outburst and the previous asymmetric outbursts might be caused by the reprocessing
   of the X-ray photons in the circumstellar disc or the tidal displacement of a large amount
   of material from the circumstellar disc or the outer layers of the donor star
   during the periastron passage of the neutron star, which led to an increase in size of the region emitting in the $I$ band.
   Further multi-wavelength
   observations are necessary to discriminate among the different scenarios proposed
   to explain the puzzling optical and X-ray properties of \src.}


   \maketitle
%

\begin{figure*}[ht!]
\centering
\includegraphics[width=17cm]{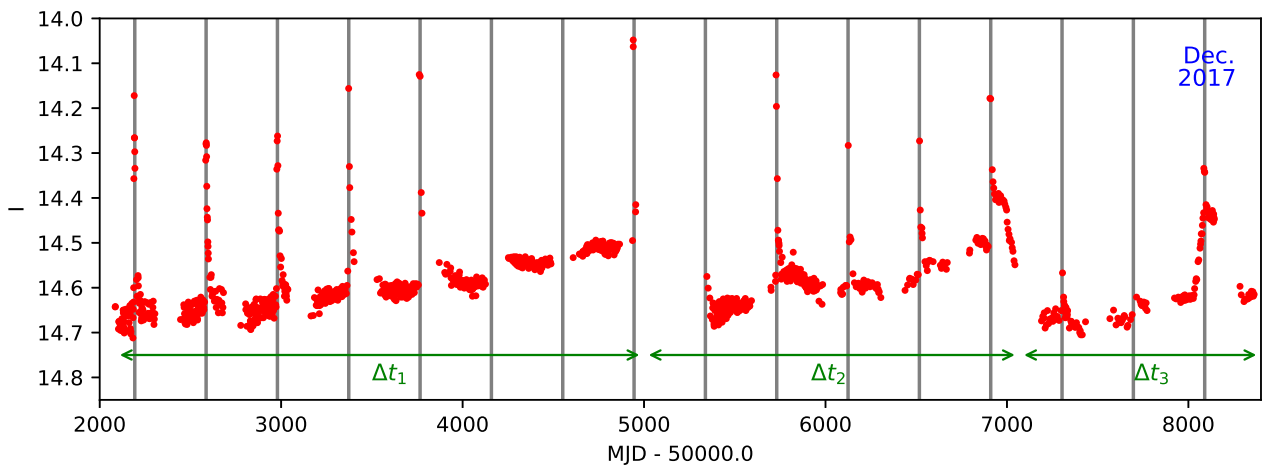}
\caption{OGLE light curve of \src\ from June 2001 to September 2018. The grey vertical lines show
the times of the peak of the optical outbursts according to the ephemeris calculated by \citet{Schmidtke13}. The green arrows at the bottom of the figure show the three epochs of gradual brightening of the basal emission (see Sect. \ref{optical photometry}).}
\label{fig:uno}
\end{figure*}

\section{Introduction} \label{sect intro}


Be/X-ray binaries (Be/XRBs, see \citealt{Reig11} for a review)
usually contain a neutron star (NS) or, rarely, a black hole 
(\citealt{Casares14} and e.g. \citealt{Brown18})
 accreting material from an early-type Oe-Be star.
Oe-Be stars are O-B type stars of luminosity class 
III-V which, at some stages in their lives, show emission lines in their spectra
(mostly in the Balmer and Paschen series) and an enhanced infrared (IR) radiation
called IR excess \citep{Waters89}.
These features in their spectra are produced in a disc of dense gas lying on the
equatorial plane of the donor star, called the circumstellar disc.
The best-studied emission line of Be stars is the H$\alpha$ line.
Its morphology and variability give fundamental information about the structure,
orientation, and evolution of the circumstellar disc (see e.g. \citealt{Hanuschik95}
and \citealt{Rivinius13} for a recent review).

The optical luminosity of Be/XRBs is dominated
by the radiation coming from the surface of the Be star
and the circumstellar disc (e.g. \citealt{vanParadijs98}).
Usually, their optical light curves show super-orbital ($\gtrsim 100$\,d) 
irregular or quasi-periodic variability, with typical amplitudes in the $I$ band
of $\Delta m_{\rm I}\lesssim 0.6$, which are believed 
to be related to the formation and depletion
of the circumstellar disc (\citealt{Rajo11}, \citealt{Reig15} and references therein). 
The variability characteristics differ from source to source.
They depend not only on the intrinsic properties of each source,
but also on the orientation of the circumstellar disc with respect
to the line of sight (e.g. \citealt{Haubois12}).
A small fraction of Be/XRBs shows orbital periodicities with
amplitudes $\Delta m_{\rm I}\approx 0.01-0.2$.
This variability is usually ascribed to perturbations of the
circumstellar disc produced by the orbiting NS.
Some Be/XRBs show a fast rise-exponential decay (FRED) profile,
which in some cases is associated with the orbital period.
\citet{Bird12} proposed using the `FRED-iness' of these profiles
as an indicator to distinguish between orbital periods and other mechanisms
like aliased radial and non-radial pulsations and long-term 
variability of the circumstellar disc.
Despite the efforts and the valuable results obtained so far, 
a detailed model to explain the mechanism producing the FRED
orbital modulation observed in some Be/XRBs is still missing.

As the name suggests, the main observational property that characterises the class
of Be/XRBs is their bright X-ray emission.
It is caused by the accretion of the dense material of the circumstellar disc
on the compact object \citep{Maraschi76}.
These episodes of high accretion are sporadic, can last several weeks, 
and can reach X-ray luminosities of $\approx 10^{39}$\,erg\,s$^{-1}$ 
(e.g. \citealt{Reig11}; \citealt{Townsend17}).
Based on observations and theoretical models, it has recently been argued
that the primary properties of the X-ray outbursts observed in most of Be/XRBs
(e.g. the point of their onset along the orbit, their duration,
the outburst rates, the X-ray luminosity 
and its variability) depend on the complicated interaction between
the compact object and the circumstellar disc, whose structure can on some occasions
deviate significantly from a symmetric geometry
\citep{Laplace17, Negueruela01, Reig07, Moritani13, Martin11, Martin14a, Martin14b, Okazaki02}.
To understand the properties of the optical and X-ray outbursts and the mechanisms which trigger them, it is of great importance to collect information about the
gas outflow from the Be star of these binary systems through 
multi-wavelength observations in X-ray and optical bands.


One of the few remarkable exceptions to the typical low-amplitude
variability observed in Be/XRBs is
\src\ (SXP\,756), which  shows relatively short ($\sim 30$ days), bright ($\Delta m_I \approx 0.5$), and periodic
outbursts ($P_{\rm orb}=393.1 \pm 0.4$\,d \citealt{Schmidtke13})
that are simultaneous with  X-ray outbursts, with a periodicity that is thought to correspond
to the orbital period of the system \citep{Cowley03, Coe04, Laycock05, Galache08}.
\src\ is located in the Small Magellanic Cloud (SMC)
and  is composed of a $\sim 750$\,s pulsar \citep{Yokogawa00} 
orbiting around a O9.5-B0.5\,III-V  star \citep{Edge03, McBride08}.
The photometric light curves based on the data from the 
MAssive Compact Halo Object Project (MACHO, \citealt{Alcock97})
and the Optical Gravitational Lensing Experiment (OGLE, Sect. \ref{sect. ogle})
of \src\ presented by \citet{Rajo11} showed that 
the amplitude of the outbursts vary and their peak luminosities increase 
with the optical emission between outbursts (hereafter basal emission).
MACHO light curves in the $V$ and $R$ bands show that \src\ becomes redder during most outbursts,
suggesting that the emitting region gets cooler \citep{Cowley03}.
The optical light curve folded at the orbital period shows
a sharp and asymmetric peak ($\Delta t \approx 30-40$\,d), making it one of the best
examples of a FRED profile \citep{Rajo11, Bird12, Coe04}.
\citet{Cowley03} found a quasi-periodic variability in the optical band of \src,
on a timescale of $\sim 11$\,d. They suggested that the observed
modulation might be associated with the rotation of the disc of the Be star.
Then \citet{Coe04} proposed that the $\sim 11$\,d periodicity might be induced
by the periastron passage of the NS which disturbs the geometry of the
circumstellar disc.
\citet{Schmidtke13} proposed that the $\sim 11$\,d signal might 
actually be an alias produced by a signal at $\sim 0.917$\,d.
The light curve folded at this latter period shows a sinusoidal modulation,
which is consistent with the expected optical variability 
due to non-radial pulsation of a Be star.

A spectroscopic study of the donor star of \src\ 
carried out during an optical outburst (on 2001 November 6-12),
showed a double-peaked H$\alpha$ emission profile,
indicating the presence of a circumstellar disc not observed face-on \citep{Edge03}.
The optical outburst was accompanied by an
X-ray outburst detected by \emph{RXTE}
\citep{Coe04,Ducci18}.

Since its discovery, \src\ has been observed in X-rays with several satellites.
As mentioned above, it shows periodic X-ray outbursts
with peak luminosities of $\approx 10^{37}$\,erg\,s$^{-1}$
which are in phase with the optical outbursts, thought to occur near  periastron \citep{Coe04, Laycock05, Galache08}.
\src\ shows high X-ray variability along the orbit, spanning more than three orders of magnitude,
with high luminosity states ($L_{\rm x} \geq 5\times 10^{35}$\,erg\,s$^{-1}$) far from periastron,
and possibly two cases of anomalous fast variability \citep{Ducci18}.
The enhanced X-ray luminosity observed far from periastron during some orbital cycles
might be due to the NS experiencing long periods of high accretion rate all along its orbit,
likely caused by an extended and warped circumstellar disc 
or some other unknown mechanism \citep{Coe04,Ducci18}.

\begin{figure}
   \centering
   \includegraphics[width=\columnwidth]{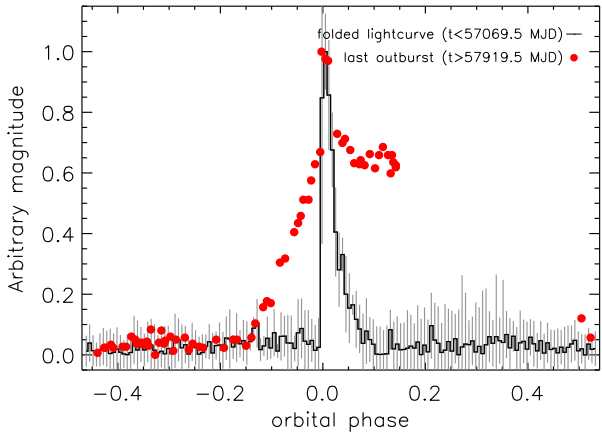}
      \caption{Black line: folded OGLE light curve of \src\ 
      obtained using data before $t \le 57069.5$\,MJD. Red points: OGLE outburst observed in December 2017. Both light curves are folded on the 393.1\,d period \citep{Schmidtke13}.}
         \label{fig folded}
   \end{figure}

The optical and X-ray variability of \src\ is unusual and still poorly understood.
Therefore, we planned a monitoring campaign
at different wavelengths, performed during the
last outburst of December 2017, to investigate it more thoroughly.
We made use of X-ray data from the Neil Gehrels Swift Observatory (\sw),
simultaneously with spectroscopic and photometric observations in optical/UV bands
with the Anglo-Australian Telescope (AAT) and \sw/UVOT, especially focused on the study of the variability of the 
H$\alpha$ line, which provided with us important information about the properties of the
circumstellar disc surrounding the system.
In addition, we used the data from the OGLE survey
collected over the last 17 years (Fig. \ref{fig folded}).
The results of these observations are presented in Sect. \ref{results}
and discussed in Sect. \ref{discussion}.

%
\begin{figure}
\centering
\includegraphics[width=\columnwidth]{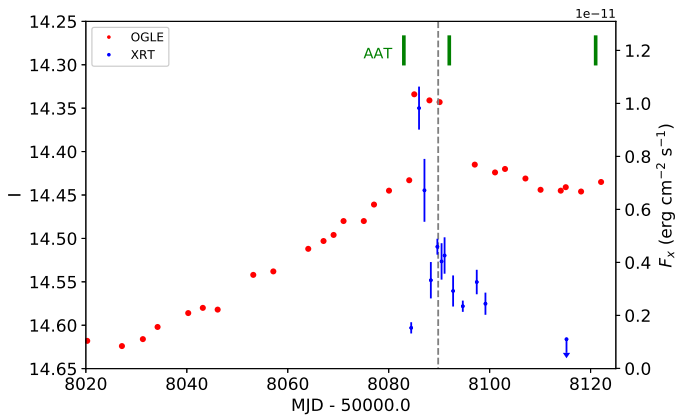}
\caption{Magnified view of the December 2017 outburst seen by OGLE (red points). 
  Superimposed are the 0.3$-$8\,keV fluxes from the \sw/XRT monitoring of \src\ (blue points), 
  and the times of the AAT observations (green bars).
  The grey vertical dashed line shows the expected time of the peak
  of the optical outburst calculated with the ephemeris of \citet{Schmidtke13}.}
\label{fig:tre}
\end{figure}

\section{Data analysis}

\subsection{\sw}

Table \ref{xrtobs} shows the log of the \sw\ \citep{Gehrels04} observations analysed in this paper.
These data were obtained as a monitoring program of 12 observations
carried out near the expected outburst, based on the ephemeris calculated by \citet{Schmidtke13}.
We processed and analysed X-ray Telescope (XRT, \citealt{Burrows05}) data using the standard software 
(FTOOLS\footnote{\url{https://heasarc.gsfc.nasa.gov/docs/software/lheasoft/ftools/}} v6.24)
and calibration (CALDB\footnote{\url{https://heasarc.gsfc.nasa.gov/docs/heasarc/caldb/caldb_intro.html}} 20180411).
XRT data were processed and filtered using {\sc xrtpipeline} (v0.13.4).
Source events were extracted from a circular region with a radius of 10 pixels (1 pixel = 2.36$^{\prime\prime}$),
while the background was extracted from an annular region centred at the source position, with an inner radius
of 30 pixels and an external radius of 70 pixels.
The data were not affected by pile-up.
The XRT light curve was corrected for point spread function losses, vignetting, and then background subctracted.
Ancillary response files were created with the task {\sc xrtmkarf}. The spectrum was extracted
from all the datasets (except for the last observation, where the source was not detected) of Table \ref{xrtobs} combined
and analysed using {\sc xspec} (v12.10.0c, \citealt{Arnaud96}).

The UV/Optical Telescope (UVOT, \citealt{Roming05}) observed \src\ simultaneously with XRT,
using six filters: ($V\approx 5468$\,\AA, $B\approx 4392$\,\AA, $U\approx 3465$\,\AA,
$w1\approx 2600$\,\AA, $m2\approx 2246$\,\AA, $w2\approx 1928$\,\AA).
The data analysis was performed using the tasks {\sc uvotimsum} and {\sc uvotsource}. 
To calculate the magnitude, we adopted a circular region with radius of $5^{\prime\prime}$ centred on the source position
and a circular region with a radius of $10^{\prime\prime}$ for the background.

\subsection{Anglo-Australian Telescope}

Three optical spectroscopic observations of \src\ were performed with the
3.9\,m Anglo-Australian Telescope (AAT) during the nights of 2017 November 26, December 5,
and 2018 January 3.
Observations were carried out with the High Efficiency and Resolution Multi-Element Spectrograph (HERMES, \citealt{Sheinis16}).
HERMES is a four-channel fibre-fed spectrograph with high resolution (R=$\lambda/\Delta\lambda\sim28000$) and multi-object capability, sensitive in the range $471.5-788.7\,$nm.
 The source was observed with a $3\times2400\,$sec of good exposure time and a $2\,$arcsec slit width.
All runs were performed with the $2785$ line/mm cross-dispersing grating.
Each channel has individual CCD cameras, each of $4k\times4k$, of the E2V family with $15\mu$m pixels, operating at about $170\,$K.
We reduced all data using a standard \texttt{2drfr} software package\footnote{\url{https://www.aao.gov.au/science/software/2dfdr}} dedicated to reducing HERMES data, and standard Image Reduction and Analysis Facility
(IRAF)\footnote{\url{iraf.noao.edu}} packages (version 2.16).
Images were bias- and flatfield-corrected, and wavelength calibration using arc lamp spectra (thorium-argon)  was applied.

\subsection{OGLE}
\label{sect. ogle}

OGLE is a photometric sky survey
that began in April 1992 \citep{Udalski15}. It monitors regularly the Galactic Bulge, part of the Galactic disc,
and the Magellanic Clouds, and it provides photometry in the $I$ band with 
uncertainties of about $5\times 10^{-3}$\,mag for the SMC fields (an analysis of the OGLE errors can be found in \citealt{Skowron16}).
During the first phase (1992--1995), the main goal of the project was the detection of gravitational microlensing events 
to constrain the nature of dark matter (e.g. \citealt{Paczynski86}).
However, thanks to the constant monitoring of about a billion  stars,
OGLE proved to be a useful project for the study of numerous other physical phenomena.
These include the study of the optical variability of Be/XRBs (\citealt{Rajo11, Bird12,Schmidtke13}).
The photometric data used in this paper have been retrieved from the OGLE\,III and IV 
X-ray variables OGLE monitoring (XROM, \citealt{Udalski08}) web sites\footnote{\url{http://ogle.astrouw.edu.pl/ogle3/xrom/xrom.html}}
\footnote{\url{http://ogle.astrouw.edu.pl/ogle4/xrom/xrom.html}}
and have been corrected to the standard $I$-band system.

 \begin{table}  
 \begin{center}         
\caption{Summary of the {\it Swift}/XRT observations.\label{xrtobs}}
\resizebox{\columnwidth}{!}{    
 \begin{tabular}{llll} 
 \hline 
 \hline 
 \noalign{\smallskip} 
  ObsID     & Start time  (UT)  & End time   (UT) & Exp.   \\ 
            &                   &                 &(s)     \\
  \noalign{\smallskip} 
 \hline 
 \noalign{\smallskip} 
00010380001     & 2017-11-27 08:23:08   &       2017-11-27 13:29:52     &       4947    \\
00010380002     & 2017-11-28 22:29:48   &       2017-11-29 01:54:21     &       1978    \\
00010380004     & 2017-11-30 01:51:00   &       2017-11-30 01:59:54     &       534     \\
00010380005     & 2017-12-01 08:09:58   &       2017-12-01 08:27:53     &       1076    \\
00010380006     & 2017-12-02 11:12:03   &       2017-12-02 19:30:52     &       7023    \\
00010380007     & 2017-12-03 11:09:51   &       2017-12-03 11:26:54     &       1023    \\
00010380008     & 2017-12-04 01:27:31   &       2017-12-04 01:45:52     &       1101    \\
00010380009     & 2017-12-05 12:23:38   &       2017-12-05 23:59:54     &       1086    \\
00010380010     & 2017-12-07 08:58:56   &       2017-12-07 23:33:53     &       6609    \\
00010380012     & 2017-12-10 05:44:59   &       2017-12-10 15:26:53     &       1667    \\
00010380013     & 2017-12-12 02:12:12   &       2017-12-12 05:37:54     &       1617    \\
00010380014     & 2017-12-28 02:49:50   &       2017-12-28 07:51:54     &       1259    \\
  \noalign{\smallskip}
  \hline
  \end{tabular}
}
  \end{center}
  \end{table}

   \begin{figure}
   \centering
   \includegraphics[angle=-90,width=\columnwidth]{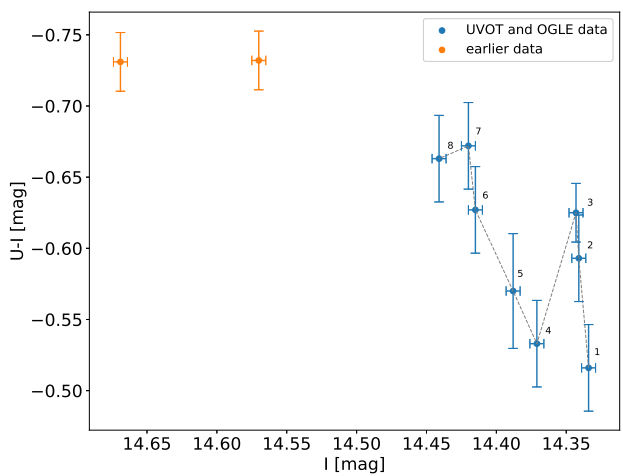}
      \caption{Colour magnitude diagram of \src\ based on UVOT and OGLE 
        data of the December 2017 outburst (blue points) and in two previous observations
        carried out far from periastron (yellow points).
        Labels from 1 to 8 show the timeline of the observations
        during the December 2017 outburst (see also Fig. \ref{xrt lcr}).}
         \label{UI_over_I}
   \end{figure}

   \begin{figure}
   \centering
   \includegraphics[width=9cm]{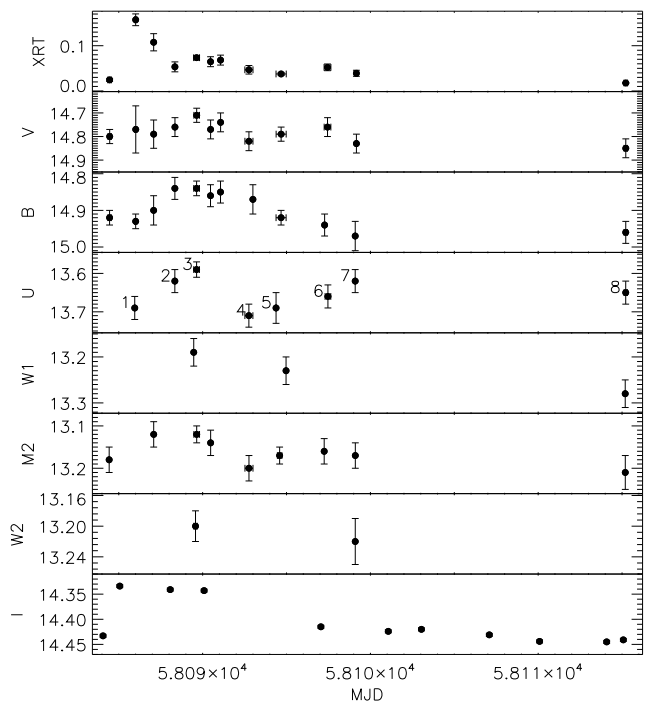}
      \caption{\sw/XRT (0.3$-$10\,keV), UVOT ($V$, $B$, $U$, $W1$, $M2$, $W2$ bands), and OGLE ($I$ band) 
        light curves of \src\ obtained during the December 2017 monitoring campaign.
        Error bars indicate the 1$\sigma$ statistical uncertainties.
        OGLE error bars and symbols are similar in  size.
        UVOT systematic errors are comparable to the statistical ones.}
         \label{xrt lcr}
   \end{figure}

\section{Results}
\label{results}

\subsection{Optical photometry}
\label{optical photometry}

Figure \ref{fig:uno} shows that the optical outbursts observed in the OGLE survey
are superimposed on three epochs of gradual brightening of the basal emission
($\Delta t_1 = 52085.4 - 55000$\,MJD; $\Delta t_2 = 55000-57069$\,MJD; 
$\Delta t_3 > 57069$\,MJD).
From the outburst at $\sim 56800$\,MJD, the 
main features characterising the optical variability of \src\
changed significantly.
The most important new pieces of information after this event emerging from Fig. \ref{fig:uno} are the following:
\begin{itemize}
\item the decay tail of the outburst observed at $\sim 56800$\,MJD
  is much longer ($\sim 140$ days) than those previously observed
  and its shape is also peculiar as it shows a hump during the decay phase;
\item after this anomalous outburst, the basal luminosity decreased to $I\approx 14.6$,
  i.e. about 0.15 magnitudes fainter than the magnitude prior to the outburst.
  Since then, and until the last outburst, \src\ showed a rebrightening of the basal emission similar to that previously observed;
\item in the two successive orbital cycles, the amplitude of the optical outbursts
  were much lower than in the past.
  The shape of the alleged outburst during the second cycle was peculiar; it was similar to a step.
  However, it is important to note that there is a gap of about 24 days in the data
  at the expected peak position;
\item The last outburst of Fig. \ref{fig:uno}, which peaks at $\sim 58090$\,MJD,
  shows a longer rise timescale (about 60 days)
  compared to the previous ones (the outbursts with more frequent sampling show
  a rise time of about 2--5 days) and also a longer decay timescale.
  The lack of data between $\sim 58142$\,MJD and $\sim 58284$\,MJD  does not allow us to estimate its duration. Despite this gap of data, we note that the decay phase
  of this outburst shows a peculiar hump, already observed in the bright outburst observed at $\sim 56800$\,MJD.
\end{itemize}

The anomalous shape of the December 2017 ($\approx$\,58090\,MJD) outburst
is highlighted in Fig. \ref{fig folded}, where it is compared to the FRED profile 
of the OGLE light curve folded on the 393.1\,d period.
We created the folded light curve from the detrended OGLE light curve obtained from two polynomial
functions (three orders each, in the ranges $\Delta t_1$ and $\Delta t_2$) fitted to the OGLE data where the periods of the outbursts were removed,
and then subtracted from the complete light curve, i.e. the one which includes the outbursts.
Figure \ref{fig:tre} shows that during the December 2017 outburst, 
the $I$-band luminosity of \src\ increased almost constantly up to $t\approx 58085$\,MJD,
where the source showed a sudden increase in the luminosity that was simultaneous (within $\sim$1 day)
with the onset of the X-ray outburst observed by \sw/XRT.

Figure \ref{UI_over_I} shows the colour-magnitude diagram $U-I$ versus $I$ obtained from the UVOT
and OGLE data of the last outburst (blue, see Fig. \ref{xrt lcr}) 
and of two previous observations
carried out far from periastron (yellow:
MJD$_1\approx55797.77$; MJD$_2=57289.16$, corresponding to the orbital
phases $\phi_1=0.17$ and $\phi_2=0.96$;
these two measurements were reported in \citealt{Ducci18}).
The $I$ magnitudes of Fig. \ref{UI_over_I} corresponding
to the $U$ points 4, 5, and 7, were obtained from interpolation.
The plot shows that the source reddens when it goes in outburst
and it cools during the outburst.
This behaviour was previously observed by \citet{Cowley03} in \src\ using $V$, $R$, and $I$ magnitudes provided
by the MACHO and OGLE-II survey,
and it was explained with a brightening of the
circumstellar disc during the outburst.

\subsection{\sw/XRT}

\begin{figure}
   \centering
   \includegraphics[bb= 75 -10 583 710, angle=-90,width=\columnwidth]{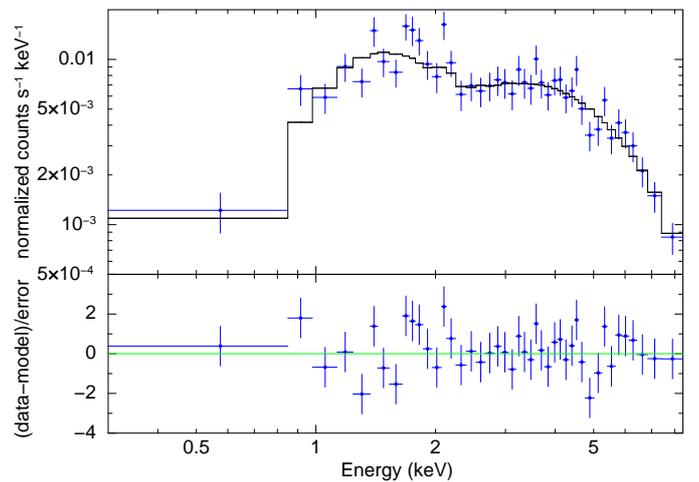}
      \caption{Average \sw/XRT spectrum of the December--January 2018 campaign
        fitted with an absorbed power law (see Table \ref{Table xrt}). The residuals of the fit are shown in the bottom panel.}
         \label{xrt spectrum}
   \end{figure}

\begin{figure}
   \centering
   \includegraphics[bb=88 371 558 700,width=\columnwidth]{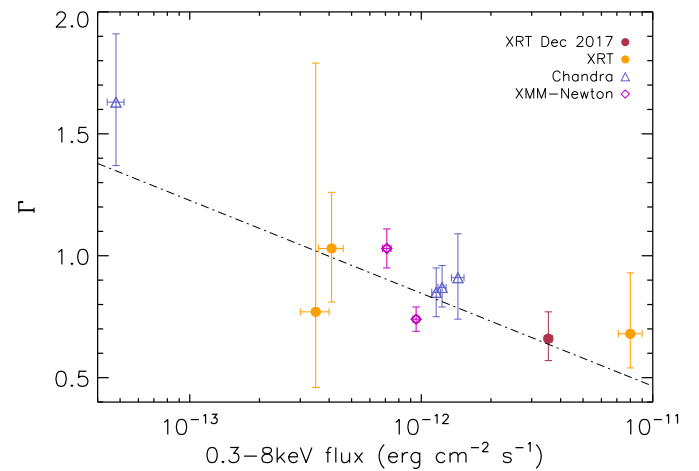}
      \caption{Photon index ($\Gamma$) as a function of the absorbed flux. Errors are quoted at the 1$\sigma$ confidence level.
        The plotted values correspond to those listed in Table 3 of \citet{Ducci18} (yellow circles, blue triangles,
        magenta diamonds) and Table \ref{Table xrt} of this paper (red circle).}
         \label{correlation}
   \end{figure}

The average XRT spectrum during the outburst can be
well fitted ($\chi^2_\nu=1.160$, 41 d.o.f.) with
an absorbed ($N_{\rm H}=4.8{+1.1\atop -0.9}\times 10^{21}$\,cm$^{-2}$;
calculated using {\tt tbabs} model in {\tt xspec}, \citealt{Wilms00})
power law ($\Gamma=0.66{+0.11 \atop -0.09}$).
The parameters of the fit are listed in Table \ref{Table xrt},
and the average spectrum is shown in Fig. \ref{xrt spectrum}
(all  errors  quoted at the 1$\sigma$ confidence level).
\citet{Ducci18} showed an anti-correlation between $\Gamma$ and the flux
using previous X-ray observations of \src.
Considering the new measure shown in Table \ref{Table xrt},
we increased the significance of this anti-correlation:
the Pearson's linear coefficient 
becomes $r=-0.85$  and the null hypothesis probability
in the $\log_{10}x-y$ space is $p=1.8$\%.
The spectral slope softens at lower fluxes following the relation
$\Gamma=(-3.72 \pm 1.3)\times \log_{10} F_{\rm x} -(0.38 \pm 0.11)$,
where the errors are quoted at the  1$\sigma$ confidence level.
The anti-correlation between $\Gamma$ and the flux including the new \sw/XRT observation
is shown in Fig. \ref{correlation}.

\subsection{H$\alpha$ emission line}
\label{sect. Halpha}

\begin{figure*}
   \centering
   \includegraphics[width=13cm]{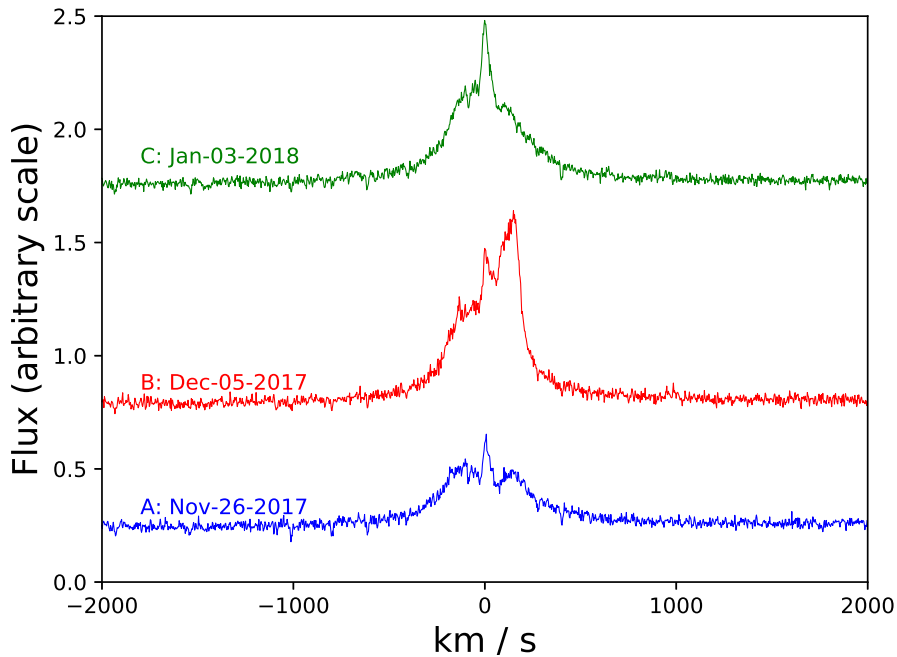}
      \caption{H$\alpha$ emission lines of \src\ during the December 2017 outburst. The velocity scale is normalised to the recession velocity
      of the SMC of $\approx 145.6$\,km\,s$^{-1}$ \citep{Harris06}.
        }
         \label{fig Halpha}
   \end{figure*}

The H$\alpha$ line was observed  three times, at the peak of the December 2017 outburst
and during the fading phase, for about 40 days.
The line profiles are shown in Fig. \ref{fig Halpha}.
Dealing with double-peaked profiles, the equivalent width (EW)  was calculated using the \textit{splot} task of
IRAF, which employs the method of direct integration of the flux across the feature.
Moreover, double-peaked profiles have also been fitted using the deblending routine
available in IRAF. 
The three peaks have been modelled  with three Voigt functions, which fitted the line wings better than Gaussian functions, returning line centres  and intensities  of the three peaks above the continuum. 
By these, we calculated the separation between the red and blue peaks $\Delta V$.
Following the methodology of \citet{Reig10} and \citet{Malacaria17}, to derive the H$\alpha$ line spectral parameters we iterated the fitting procedure twelve times for each double-peaked feature and sampled, at each iteration, a slightly different point of the continuum (whose definition is the main source of uncertainty during this process).
Final values and errors of line parameters were calculated as the average and standard deviation over those iterated measures.
The equivalent width $EW$ and the separation between
red and blue peaks $\Delta V$ of each observation
are given in Table \ref{tab line}.
The H$\alpha$ line
showed a high variability and a peculiar shape
in all three observations.
In particular, during observation A, carried out slightly before the
peak of the outburst (see Fig. \ref{fig:uno}),
the H$\alpha$ profile showed a three-peak profile with
a narrow central peak and two side peaks of about the same
height ($V=R$).
Nine days later (observation B), when the optical emission from the source was fading, 
the H$\alpha$ profile changed significantly:
its intensity increased, and the height of the red peak became
higher than the blue peak ($V<R$).
The observed $V/R$ variability could be due
to density perturbations in the circumstellar disc caused by the NS
or to one-armed disc oscillations of the circumstellar disc similar to
those observed in other Be stars (see e.g. \citealt{Rivinius13} and references therein).
In observation C, carried out about one month later, when \src\
was much fainter in the $I$ band and not detected in X-ray,
the H$\alpha$ profile still shows a relatively high intensity compared
to observation B and especially observation A.
The height of the blue and red peaks was much lower compared to the central peak
(they almost disappeared).

\begin{table}
\begin{center}
\caption{Best fit parameters of the \sw/XRT average spectrum.}
\label{Table xrt}
\begin{tabular}{lc}
\hline
\hline
\noalign{\smallskip}
Parameter                             &     Value   \\
\hline
\noalign{\smallskip}
$N_{\rm H}$ $(10^{21})$\,cm$^{-2}$        &  $4.8{+1.1\atop -0.9}$   \\
\noalign{\smallskip}
$\Gamma$                               &  $0.66{+0.11 \atop -0.09}$    \\
\noalign{\smallskip}
$\chi^2_\nu$  (d.o.f.)                  &  1.160 (41)     \\ 
\noalign{\smallskip}
absorbed flux                          &  $3.55 \pm 0.14 \times 10^{-12}$ \\
\noalign{\smallskip}
unabsorbed flux                        &  $3.97{+0.15 \atop -0.13} \times 10^{-12}$   \\
\noalign{\smallskip}
\hline
\end{tabular}
\end{center}
{\small Notes: fluxes are in units of erg\,cm$^{-2}$\,s$^{-1}$ in the 0.3$-$8\,keV energy range. Errors are quoted at the $1\sigma$ confidence level.}
\end{table}

\begin{table}   
 \begin{center}         
\caption{Spectral parameters of the H$\alpha$ emission line. 
Errors quoted at the $1\sigma$ confidence level. \label{tab line}}
 \begin{tabular}{llcc} 
 \hline 
 \hline 
 \noalign{\smallskip}     
 Obs. & Date  &       $EW$       &    $\Delta V$        \\
      &       &       [\AA]      &  [km\,s$^{-1}$]       \\
  \noalign{\smallskip} 
 \hline 
 \noalign{\smallskip} 
A & 2017-Nov-26  &   $-17.8 \pm 3.0$  &   $280.3 \pm 20.9$     \\
   \noalign{\smallskip}
B & 2017-Dec-05  &   $-24.3 \pm 4.1$  &  $262.1 \pm 4.5$      \\
     \noalign{\smallskip}
C & 2018-Jan-03  &   $-20.8 \pm 3.9$  &  $220.3 \pm 3.3$      \\
  \noalign{\smallskip}
  \hline
  \end{tabular}
  \end{center}
  \end{table}

\section{Discussion}
\label{discussion}

In Sect. \ref{optical photometry} we showed that \src\
experienced a change in its optical properties after the outburst
observed at $\sim 56800$\,MJD. The decrease in the $I$ magnitude
after this outburst and the low amplitude of the two subsequent
outbursts compared to the previous ones displayed by the system,
might be due to a rapid depletion of the circumstellar disc
(likely triggered by the tidal interaction with the NS)
which then slowly formed again.
Unfortunately, the binary system was not observed at other wavelengths
during the anomalous outburst of $\sim 56800$\,MJD and 
the lack of H$\alpha$ observations
during and immediately after that event prevents us from testing our hypothesis.

The multi-wavelength monitoring campaign of \src\ carried out in December 2017
revealed an anomalous outburst whose optical emission was characterised by
rise and decay timescales longer than in the previous outbursts.
As shown in Fig. \ref{fig folded}, the outburst did not show
the distinctive FRED profile observed in the past.
The rise of the outburst can be divided in two parts.
In the first part the flux increases slowly,
starting from $t_0 \approx 58020$\,MJD
(corresponding to the phase $\phi \approx -0.15$ in Fig. \ref{fig folded}) and rising  to $t_1=58085$\,MJD. 
The rise timescale of this part of the outburst is relatively fast compared
to the typical timescales of the formation 
of circumstellar discs in Be/XRBs (see e.g. \citealt{Haubois12}),
but it is roughly consistent with those seen in some of them (e.g. \citealt{Kiziloglu09}; \citealt{Rajo11}).
We suggest that it might be due to an increment in the size
of the circumstellar disc
triggered by the tidal interactions of a NS in an eccentric orbit that approaches  periastron.

In the second part of the light curve, from $t_1$ to the peak, 
\src\ displays 
a steep jump in the optical brightness 
synchronised with the onset of the
X-ray outburst.
The fast increase in luminosity at $t_1$ is similar to that of the FRED 
outbursts observed prior to $\approx 56800$\,MJD, though with much lower amplitude,
and might suggest that another mechanism (likely triggered by the X-ray outburst)
could be the cause of the bright peak of this outburst and also the previous 
asymmetric outbursts.

The  current explanation for the FRED-shaped optical outbursts
simultaneous with the X-ray outbursts was proposed by
\citet{Cowley03} and \citet{Coe04}. They suggested that the optical outbursts
were caused by the disruption of the circumstellar disc
caused by its interaction with the NS.
They also pointed out that a similar outburst behaviour
was observed in another Be/XRB, A0538$-$66.
In this system, an optical variability spanning more than 2 magnitudes was observed
simultaneously with bright X-ray outbursts ($L_{\rm x}\approx 10^{39}$\,erg\,s$^{-1}$,
\citealt{Charles83} and references therein).
\citet{Charles83}, \citet{Maraschi83}, and \citet{Apparao88} suggested that the bright optical outbursts
of A0538$-$66 were produced by the reprocessing of the X-ray photons emitted
by the NS in the circumstellar envelope around the binary system.
A similar mechanism could explain the asymmetric outbursts observed
in \src\ prior to $\approx 56800$\,MJD and the sudden increase in  luminosity
at $t_1$ observed in the December 2017 outburst.
A similar possibility was also considered by \citet{McGowan08}
for the optical variability shown by SXP\,46.6,
an accreting pulsar in the SMC.
Alternatively, 
these fast outbursts might be caused
by the tidal displacement of a large amount of material 
from the circumstellar disc or the outer layers of the donor star
during the periastron passage of the NS, which then translates to a significant
increase in the size of the emitting region and hence in the 
optical luminosity.

In Sect. \ref{sect. Halpha} we presented the results of three observations
of the H$\alpha$ line carried out during the December 2017 outburst.
Previously, the H$\alpha$ profile of \src\ was observed only once by \citet{Edge03},
during another optical outburst (at the orbital phase $\phi \approx 0.07$, according to the ephemeris of \citealt{Schmidtke13}). 
In that observation it showed the typical
double peak seen in many other Be stars with a circumstellar disc observed at a high inclination angle.
In our observations, the H$\alpha$ profile showed three peaks
with variable reciprocal intensity.
Although with remarkably different properties  from those of \src,
only a few other Be/XRBs 
showed anomalous H$\alpha$ profiles. Among these,
two interesting cases are
4U\,0115+63 and A0535+26.
\citet{Negueruela01} reported the observations of a succession of single-peaked
and shell profiles in the emission lines of 4U\,0115+63
on a timescale of several months.
Shell profiles are usually associated with Be discs seen edge-on,
while single-peaked profiles are observed in pole-on Be discs.
They interpreted their observations with a highly tilted warped
precessing circumstellar disc. The large variability of the H$\alpha$ emission line
of 4U\,0115+63 was associated with the onset of giant X-ray outbursts.
\citet{Moritani11, Moritani13}, studied the H$\alpha$ line profile variability
of A0535+26 from 2009 to 2012. During this period, the source displayed several X-ray outbursts.
In 2009, during a giant X-ray outburst, the line profile showed three peaks.
They interpreted these features as the result of a precessing warped
circumstellar disc.
Each of the three peaks was produced in different regions of the warped disc
observed from different angles.
We suggest that both the peculiar shape and the high variability of the H$\alpha$ profile
of \src\ presented in this work might be interpreted 
with a scenario similar to that proposed by \citet{Moritani11, Moritani13} for A0535+26,
namely with perturbations in the circumstellar disc driven by radiation
or by tidal interactions with the NS in a high eccentricity orbit during its periastron passage.
Similar to what was done by \citet{Moritani11},
we used the calculations presented by \citet{Porter98} to estimate
the timescale associated with the precession of a radiatively induced
warped disc around a Be star.
We found that a radiatively
driven warped disc around \src\ would have a precessing timescale
ranging from one to five months. The lower limit of about 30 days,
obtained assuming the stellar properties of a O9.5\,V star, is compatible
with the observed timescale. 
Another way to warp the disc is through tidal interactions with the NS.
\citet{Martin11, Martin14a, Martin14b} showed, with numerical simulations,
that if the circumstellar disc of a Be star and the orbital plane of the NS
are misaligned, the tidal torque exerted by the NS on the disc
can warp it and make it more elongated.
Interestingly, an extended and warped circumstellar disc,
slightly misaligned with respect to the orbital plane,
would also explain the high X-ray variability of \src\ observed
far from periastron reported in \citet{Ducci18}.\\
\indent An alternative explanation
is that the complicated shape of the H$\alpha$
line of \src\ is caused by the superposition of two contributions from the
circumstellar disc and from an accretion disc around the NS, with different reciprocal orientations.
\citet{Kiziloglu09} suggested that accretion discs around a NS
in a Be/XRB might contribute to the H$\alpha$ line.
This line has been used 
to reveal the presence of an accretion disc in other 
X-ray binary systems such as 
SS\,433 \citep{Bowler10}, IGR\,J17329$-$2731 \citep{Bozzo18},
and IGR\,J00291+5934 \citep{Lewis10}.
It is important to note that in this framework some problems would emerge for the case of \src\ that require further clarification. The central peak was not present during a previous bright X-ray and optical outburst (November 2001, see Sect. \ref{sect intro}), when the H$\alpha$ line was observed to have the typical double-peak profile produced by the circumstellar discs seen in the spectra of many Be stars. The absence of the central emission line in the previous outburst
would suggest that this component is caused by a transient accretion disc.
It is worth noting that the central peak of the H$\alpha$ line observed during the December 2017 outburst has an EW of about $-0.9\,\AA$, which is strikingly similar to that observed for the single-peaked H$\alpha$ line from the face-on accretion disc of the X-ray binary  MAXI\,J1836$-$194 \citep{Russel14}.
Although it is an attractive scenario, due to the lack of more detailed observations of \src\ it is not possible to go beyond these speculations.

\section{Conclusions}

We presented the results from a multi-wavelength monitoring campaign of 
the Be/XRB \src\ during the outburst of December 2017
in  X-ray and optical.
We observed the binary system during an anomalous optical outburst
characterised by longer rise and decay timescales
than in the previous outbursts, and without the characteristic FRED profile
observed previously.
Near the peak of the outburst, the source showed a sudden increase in the 
flux in X-ray and $I$ bands.
We proposed that the properties of the H$\alpha$ line and the
 variability of \src\ in the $I$ band indicate the presence of a highly 
perturbed circumstellar disc, likely warped and elongated
by the tidal interaction with a NS in a high eccentricity orbit during
the periastron passage.
We also suggested that the X-ray reprocessing of X-ray photons in the circumstellar
disc might be the cause of the asymmetric outbursts observed prior to $\approx 56800$\,MJD
and the jump in optical luminosity at the peak of the December 2017 outburst.
We proposed an alternative scenario, where also an accretion disc around the NS contributes to the complicated shape of the H$\alpha$ line.
Although the monitoring campaign of the December 2017 outburst
has revealed new interesting properties of \src,
the information collected so far are insufficient to draw
any firm conclusion about the peculiar optical and X-ray variability of \src.
Therefore, new multi-wavelength observations and 
the determination of the orbital parameters of the system,
such as the eccentricity and the inclination angle between 
the orbital plane and the circumstellar disc,
will be fundamental to answer some of the most important open questions about 
the puzzling properties of \src, with particular emphasis on
the morphology of the gas envelope around \src\ 
and the mechanism responsible for the simultaneous `FRED' optical
and  X-ray outbursts.

\begin{acknowledgements}
We thank the anonymous referee for the constructive comments that
helped to improve the paper.
This paper is based in part on data from the Neil Gehrels Swift Observatory
and on data acquired through the Australian Astronomical Observatory. 
We acknowledge the traditional owners of the land on which the AAT stands, 
the Gamilaraay people, and pay our respects to elders past and present.
This work is supported by the Bundesministerium f\"ur
Wirtschaft und Technologie through the Deutsches Zentrum f\"ur Luft
und Raumfahrt (grant FKZ 50 OG 1602).
P.R. acknowledges contract ASI-INAF I/004/11/0.
C.M. is supported by an appointment to the NASA Postdoctoral Program at the Marshall Space Flight Center, administered by Universities Space Research Association under contract with NASA.
The OGLE project has received funding from the National Science Centre,
Poland, grant MAESTRO 2014/14/A/ST9/00121 to AU.
J.L. is grateful for the support from the Chinese NSFC 11733009.
\end{acknowledgements}

\bibliographystyle{aa} 
\bibliography{ax0049}

\end{document}